\pdfoutput=1
\documentclass[11pt]{article}

\usepackage{naacl2021}

\usepackage{times}
\usepackage{latexsym}

\usepackage[T1]{fontenc}
\usepackage[utf8]{inputenc}

\usepackage{microtype}

\usepackage{tikz}
\usetikzlibrary{patterns}
\usetikzlibrary{shapes.callouts}
\usepackage{amssymb}
\usepackage{enumitem}
\usepackage[utf8]{inputenc}
\usepackage{url}
\usepackage{todonotes}

\pgfkeys{%
    /calloutquote/.cd,
    width/.code                   =  {\def\calloutquotewidth{#1}},
    position/.code                =  {\def\calloutquotepos{#1}}, 
    author/.code                  =  {\def\calloutquoteauthor{#1}},
    /calloutquote/.unknown/.code   =  {\let\searchname=\pgfkeyscurrentname
                                 \pgfkeysalso{\searchname/.try=#1,                                
    /tikz/\searchname/.retry=#1},\pgfkeysalso{\searchname/.try=#1,
                                  /pgf/\searchname/.retry=#1}}
                            }

\newcommand\calloutquote[2][]{%
       \pgfkeys{/calloutquote/.cd,
         width               = 5cm,
         position            = {(0,-1)},
         author              = {}}
  \pgfqkeys{/calloutquote}{#1}                   
  \node [rectangle callout,callout relative pointer={\calloutquotepos},text width=\calloutquotewidth,/calloutquote/.cd,
     #1] (tmpcall) at (0,0) {#2};
  \node at (tmpcall.pointer){\calloutquoteauthor};    
}  

\newcommand{\roundpic}[4][]{
  \tikz\node [circle, minimum width = #2,
    path picture = {
      \node [#1] at (path picture bounding box.center) {
        \includegraphics[width=#3]{#4}};
    }] {};}

\newcommand{\sep}{\Large\slash\kern-.26em\raisebox{.5ex}{*}\ \normalsize}

\title{The Online Pivot: Lessons Learned from Teaching a Text and Data Mining Course in Lockdown, Enhancing online Teaching with Pair Programming and Digital Badges}

\author{Beatrice Alex \\
Literatures, Languages and Cultures \\
University of Edinburgh \\
Edinburgh \\
Scotland, UK\\
  \texttt{balex@ed.ac.uk} \\\And
  Clare Llewellyn \\
Social and Political Science\\
University of Edinburgh \\
Edinburgh \\
Scotland, UK\\
  \texttt{clare.llewellyn@ed.ac.uk} \\
\AND

  Pawel Michal Orzechowski ~~~~~~~~~~ Maria Boutchkova \\
University of Edinburgh Business School\\
University of Edinburgh \\
Scotland, UK\\
  \texttt{Pawel.Orzechowski@ed.ac.uk Maria.Boutchkova@ed.ac.uk}}
\begin{document}
\maketitle
\begin{abstract}

In this paper we provide an account of how we ported a text and data mining course online in summer 2020 as a result of the COVID-19 pandemic and how we improved it in a second pilot run.  We describe the course, how we adapted it over the two pilot runs and what teaching techniques we used to improve students' learning and community building online.  We also provide information on the relentless feedback collected during the course which helped us to adapt our teaching from one session to the next and one pilot to the next.  We discuss the lessons learned and promote the use of innovative teaching techniques applied to the digital such as digital badges and pair programming in break-out rooms for teaching Natural Language Processing courses to beginners and students with different backgrounds.

\end{abstract}

\section{Introduction}

It was spring 2020 and it felt like we were in crisis mode. We wanted to teach a text and data mining (TDM) pilot course but because of social distancing measures we could not do it in a physical classroom. We had to learn new ways of interacting online and using a multitude of different technologies and we needed to do it fast. We had been planning this course for a while before Covid-19 hit. We were designing a TDM for Humanities and Social Science students but because of the situation we had to adapt the way we delivered it. Rather than hybrid teaching as intended, accommodating in-classroom, online synchronous and online asynchronous students, we had to fully commit to online methods in a matter of a few weeks. We decided to plunge headlong into the digital teaching world.

The easiest way would have been to post videos of a traditional style lectures – it is very tempting to take this approach. We felt, however, that it was important that we maintained what is good about teaching when everyone is in the same room, the collaboration, its social aspects, the feedback, all of which you lose when a student sits on their own in a room watching a pre-recorded lecture. 

We decided to run a TDM boot camp to virtually test our new course which we were planning as part of the Edinburgh Futures Institute (EFI) postgraduate programme.\footnote{EFI is a new institute at the University of Edinburgh which will support interdisciplinary research and teaching for the whole institution. \url{https://efi.ed.ac.uk}} We wanted to not only teach fundamental methods for text mining corpora to programming novices but also teach ourselves how to become better practitioners in teaching in an online world.

In this paper, we will describe our methods and experience for porting an in-person TDM course into the online world. In the next section we will present related publications on teaching Natural Language Processing or TDM courses.  We then describe the academic backgrounds of the teaching team (Section~\ref{section:team}) and provide an overview of our course (Section~\ref{section:course}). Sections~\ref{section:bootcamp1} and \ref{section:bootcamp2} explain how we taught and adapted it in two online pilot runs delivered in June and September 2020.  We provide information on how we collected relentless feedback during and after each course and include a detailed account of one participant of the first pilot and how it has affected her teaching (Section~\ref{section:feedback}). Finally, we summarise what we learned from these experiences (Section~\ref{section:lessons}) and lay out future plans for our TDM course (Section~\ref{section:summary-and-future}).

%the Edinburgh Futures Institute postgraduate Futures Programmes

\section{Related Work}\label{section:background}
There are two aspects that we consider of importance in relation to this work, the course content, Natural Language Processing (NLP), and the environment, teaching online during a pandemic. In this section we explore both topics.

NLP educators choose which aspects to teach based on multiple constraints such as class length, student experience, recent advancements, program focus, and even personal interest. 

Our TDM course is fundamentally designed to be cross-disciplinary as we are teaching NLP and coding to students from multiple schools and backgrounds including linguistics, social sciences and business. \citet{jurgens2018look} point out that NLP courses are designed to reflect, amongst other things, the background and experience of the students. \citet{agarwal-2013-teaching} explains that in courses such as these the majority of students, who he calls “newbies in Computer Science”, have never programmed before. He highlights that we can increase experience through homework tasks which we did both before the course and in-between each session. \citet{hearst2005teaching} states that in these circumstances it is not important to place too much emphasis on the theoretical underpinnings of NLP but to focus on providing instructions for students on what is possible and how they can use it on their own in the future. We based our approach on using the NLTK\footnote{\url{https://www.nltk.org}} and spaCy\footnote{\url{https://spacy.io}} Python libraries as well as used examples inspired by \citet{bird2009natural,bird2005nltk}. We aim to explain how text analysis works step-by-step using clear and simple examples. We thereby aspire to develop and broaden humanities and social science students' data-driven training and give them an understanding of how things work inside the box, something for which there is still a significant need in their core disciplines \cite{mcgillivray2020challenges}.

Teaching text analysis to non-computer scientists has been explored in texts such as \citet{hovy2020text}. For our course we had to consider the variety of backgrounds and experiences that this would encompass and needed to use a pre-course learning task and office hours to provide a more level knowledge starting point. We also had to design the course to keep more advanced students engaged while not intimidating learners who may find it more challenging. We used core material to explain principle concepts (such as tokens, tokenisation, and part-of-speech (POS) tagging etc.) but with a hands-on approach.  We avoided too much technical detail and put the material in the context of projects we have worked on ourselves to demonstrate how each analysis step becomes useful in practice.

As we taught our TDM course online in the context of a worldwide pandemic, we also report on related work in the area of online teaching, and with respect to the challenges in which we are teaching. Massive open online courses (MOOC) generally focus on providing online access to learning resources to a large number and wide range of participants. This has led to a desire to automate teaching and innovate digital interaction techniques in order to engage with large numbers of students. Whilst our intention was to teach a limited number of students, we hoped to use and draw upon innovation in this area in order to improve the experience for our students. E-learning and technology should not be seen as an attempt to replace or automate human teaching, although this can often be a fear articulated by teachers. In a discussion of automation within teaching \citet{bayne2015teacherbot} argues that we can design online teaching and still place human communication at the centre with technology enhancing the learning of the student. Bayne suggests that the human teacher, the student and the technology can be intertwined. We asked students to engage with digital objects and the technology to enhance their learning journey. As teachers we do not merely support the digital learner but we remain at the centre of teaching the course. 

\citet{fawns2019online} point out that online learning is a key growth area in higher education, which is even more true since the pandemic started, but that it is harder to form relationships in online courses. Therefore, we saw it as important to develop online dialogue between students in order to form communities which can improve these relationships. Building a community online can be harder but it is possible. We tried to achieve this through using a combination of traditional learning such as lectures and task-based learning such as pair programming exercises. Online learning tends to be interrupted as we are in our homes or elsewhere and have responsibilities that can take us away from the online space, bandwidth issues, dropping children at school, flatmates interrupting, phone calls, even the door bell ringing. Our teaching practices needed to be accepting of and adapted to this context.

\citet{ross2013making} discuss the issues of presence and distance in online learning. Interruptions in students' concentration are a common event when learning online and we must use resilience strategies to maintain a ‘nearness’ to our students. This includes recognising that these events are normal and that engaging is an effort, identifying affinities and creating a socialness, valuing that distraction can change our perspective and this is helpful and designing openings, events that allow and encourage student to come together and engage. Whilst designing the course we kept these ideas in focus in order to allow us to develop and enhance our online relationships and our students' learning.

\section{A Virtual Learning Experience}\label{section:virtual}

\subsection{The Team}\label{section:team}
Our team is made up of three early career academics at the University of Edinburgh. Two teaching fellows have a background in Natural Language Processing with PhDs in Computational Linguistics. The third teaching fellow has a PhD in Computer Science and frequently teaches programming to different types of audiences, including business students as well as students outside of higher education.  The author list of this paper also includes a fourth (last) author who was a participant of our first pilot, is a lecturer herself, and who has provided us with useful feedback for future iterations of this course (see Section~\ref{section:feedback-maria}).

%The author list of this paper also includes a forth author who was a participant of our first pilot course and who has provided us with useful feedback for future iterations of this course.

\subsection{Course Overview}\label{section:course}

In our data-driven society, it is increasingly essential for people throughout the private, public and third sectors to know how to analyse the wealth of information society creates each day. Our TDM course gives participants who have no or very limited coding experience the tools they need to interrogate data. This course is designed to teach non-coders how to analyse textual data using Python as the main programming language. It takes them through the required steps needed to be able to analyse and visualise information in large sets of textual document collections, or corpora.

\begin{figure*}[t]
\centering
\begin{tikzpicture}

\node[inner sep=0pt] at (0,0) {\includegraphics[width=.8\textwidth]{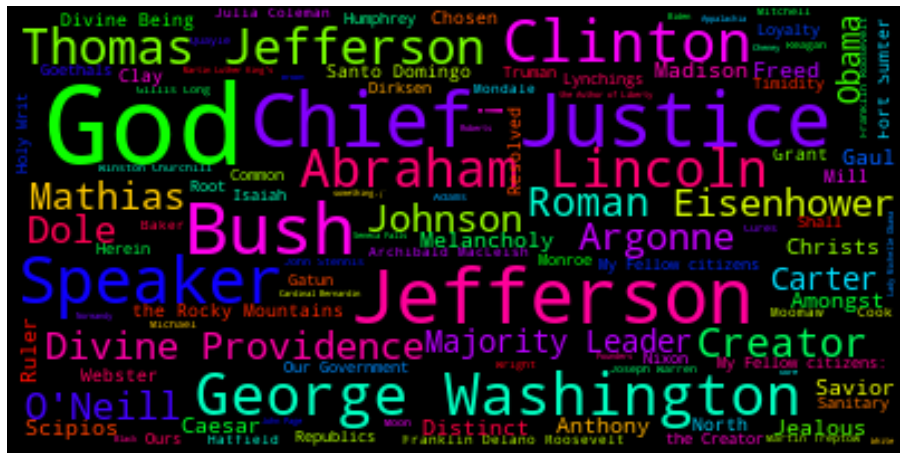}}; 

\node[inner sep=0pt] (fig) at (4,-5)
    {\includegraphics[width=.68\textwidth]{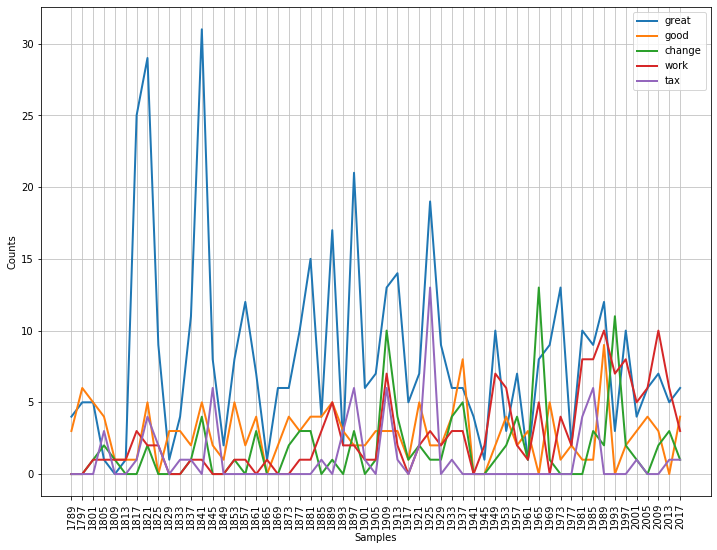}
};
\end{tikzpicture}
  \caption{Visualisations of text explorations created by the students.}
  \label{fig:data}
\end{figure*}

The course takes place over three three-hour sessions and each session introduces participants to a new topic through a short lecture. The topics build on the previous sessions and at the end of each session there is time for discussion and feedback. In the first session we start with Python for reading in and processing text and teach how individual documents are loaded and tokenised. We work with plain text files but do raise the issue that textual data can be stored in different formats. However, to keep things simple we do not cover other formats in detail in the practical sessions.

In the second session we show how this is done using much larger sets of text and add in visualisations. We used two data sets as examples, the  Medical History of British India \cite{nls-mhbi2019} made available by the National Library of Scotland\footnote{\url{https://data.nls.uk/data/digitised-collections/a-medical-history-of-british-india/}} and the inaugural addresses of all American Presidents from 1789 to 2017. We show how participants can create concordance lists, token frequency distributions in a corpus and over time as well as lexical dispersion plots and how they can perform regular expression searches using Python.  In this session we also explain that textual data can be messy and that a lot of time can be spent on cleaning and preparing data in a way that is most useful for further analysis. For example, we point students at stop words and punctuation in the results and explain how to filter them when creating frequency-based visualisations.

During the third session we cover POS-tagging and named entity recognition. This last session concludes with a lesson on visualisations of text and derived data by means of text highlighting, frequency graphs, word clouds and networks (see some examples in Figure~\ref{fig:data}). The underlying NLP tools used for this course are NLTK 3 and spaCy which are widely use for NLP research and development.  This is also where we put some of the course material in context of our own research to show how it can be applied in practice in a real project.  For example, we mentioned our previous work on collecting topic-specific Twitter datasets for further analysis \cite{llewellyn2015extracting}, on geoparsing historical and literary text \cite{clifford2016geoparsing,alex2019geoparsing} and on named entity recognition for radiology reports \cite{alex2019text,gorinski2019named}.

In the two pilots, we ran this course over three afternoon sessions on Monday, Wednesday and Friday, with an office hour on the days in-between to sort out any potential technical issues and answer questions. The main learning outcome is that by the end of the course the participants will have acquired initial TDM skills which they can use in their own research and build on by taking more advanced NLP courses or tutorials.  A main goal of this course is to teach the material in a clear step-by-step way so all Python code and the examples are specific to each task but do not go in-depth into complicated programming concepts which we believe would confuse complete novices.

\subsection{Pilot 1}\label{section:bootcamp1}

In the first pilot we wanted to test the content of this course but also different methods for teaching online. We are all likely to be teaching virtually more often in the future even once the pandemic subsides. For example, EFI was planning to run hybrid courses to students across the world, even prior to COVID-19.  In this new world, we believe that online and hybrid teaching is here to stay alongside teaching students in the classroom. Higher education will need to determine their offer of different experiences to students be they on site or participating online synchronously or asynchronously.

We limited the first pilot to 25 participants. The backgrounds of students who signed up for our course were mixed coming from Law, Linguistics and Business. Everyone was either a student or a member of staff at the University of Edinburgh, where we had advertised the course, including every level from professor to undergraduate, joining from around the world. Some students even participated from different time zones.

On each day we started with a short presentation discussing the TDM theory of what was being taught in the practical session that followed. In the first pilot this was a live lecture, not recorded, allowing us to adapt the content to questions that came up during the course. When one teacher spoke the other two managed the video chat, answering questions or dealing with specific problems from students, and raising questions to the speaker. This was something we found was essential as it was very easy to lose flow and get distracted without this help. We learned then that it would have been extremely challenging to teach this course live online single-handedly and after each session expressed appreciation that there were three of us helping each other.

We used a variety of technologies provided by the university. Learn,\footnote{\url{https://www.learn.ed.ac.uk}} our in-house virtual learning environment (VLE), was used to provide access to course materials. We met with students virtually using the Blackboard Collaborate software\footnote{\url{https://help.blackboard.com/Learn/Instructor/Interact/Blackboard_Collaborate}} which is accessible through Learn. Aside from the video itself, we used text chat, the virtual whiteboard, polls, the ability to raise a hand, breakout groups, file sharing, and screen sharing, all functionalities which have become second nature after a year of pandemic but which when we ran the first pilot were for the most part still fairly unfamiliar to many participants. We also used Noteable,\footnote{\url{https://noteable.edina.ac.uk}} the University of Edinburgh's in-house notebook platform, to provide a virtual programming environment (VPE) with Jupyter Notebooks,\footnote{\url{https://jupyter.org}} and used GitHub\footnote{\url{https://github.com}} to provide students access to the course material and code. We note that the students did not have to learn how to use GitHub, which would be a big ask for coding novices, but merely had to paste the GitHub link of the corresponding material into Noteable which then automatically loaded the material in the form of a notebook.

%X = Learn
%\footnote{\url{https://www.ed.ac.uk/information-services/learning-technology/communication}}
%Y = Noteable \footnote{\url{https://noteable.edina.ac.uk}}

Each day the students were given two sets of worked through problems using the VPE which they used directly through the VPL in their own browser. We found this to be a really important tool for everyone as it reduced the need for students to download and set up software on different operating systems and alleviated us from doing a lot of technical support to get students set up and running for all the practical parts of the course.

During the sessions the students were given a link to a GitHub repository from which they could pull new notebooks onto the VPE at the beginning of each session. The notebooks include a combination of explanations, code to run and mini or extended programming tasks. For each approximately hour-long coding session students were assigned a random buddy and which they were put in a breakout room within the Collaborate video call. By now we are used to teaching and/or learning online and have likely experienced joining break-out rooms but at the time when we ran the first pilot most of our participants had never been in a break-out room before. So that experience took some getting used to. We described it as \textit{feeling like being put in a separate room with your buddy. You can chat and share screens without being overheard by other people.} If the students got stuck on a particular coding problem or line of code and could not solve the issue together, they could raise a virtual hand and an instructor would drop into the room to help and answer questions or resolve programming issues. We also regularly popped into the rooms to see how everyone was doing, something which was well received by the students.

One of our team members is a strong proponent of pair programming \cite{williams2000strengthening,hanks2011pair}, where two students work together on a single machine to solve problems. This allows each pair of students to learn from each other as well as from their teacher(s) and thereby helps to broaden participation and to dispel the myth that programmers work on their own \cite{williams2006debunking}. We wanted to see if it was possible to take this approach into a virtual teaching environment. In addition to the students learning TDM skills, it also provided an opportunity for social interaction which was particularly welcome when we first piloted our course at the tail end of the first wave of COVID-19 in the UK and after weeks of strict lockdown with no or little opportunity to meet and interact with people outside one's own household.

One advantage of Blackboard Collaborate is that instructors are able to see visually when the people in break-out rooms are chatting to each other.  This helped us to gauge if students embraced our pair programming experiment or if they preferred to work quietly "side-by-side" but connected virtually.

After each practical session we pulled everyone back into the shared room and asked participants to fill in a quick survey to give us feedback. We answered any questions, had a quick break, and then moved onto the next notebook with a new buddy. We wrapped up each session with a short Q\&A and another round of feedback.

\subsection{Pilot 2}\label{section:bootcamp2}

By the time of the second pilot in September 2020, we had gotten a lot more used to online meetings and two members of the teaching team had trained in a summer course on hybrid teaching called An Edinburgh Model for Teaching Online.  This time we allowed 30 participants to sign up with over half of them from Scottish Government and the commercial sector, alongside university students and staff.

%and alongside university students and staff, we also invited participants from the Scottish Government and the commercial sector.

\begin{figure}[t]
\centering
\includegraphics[width=8cm]{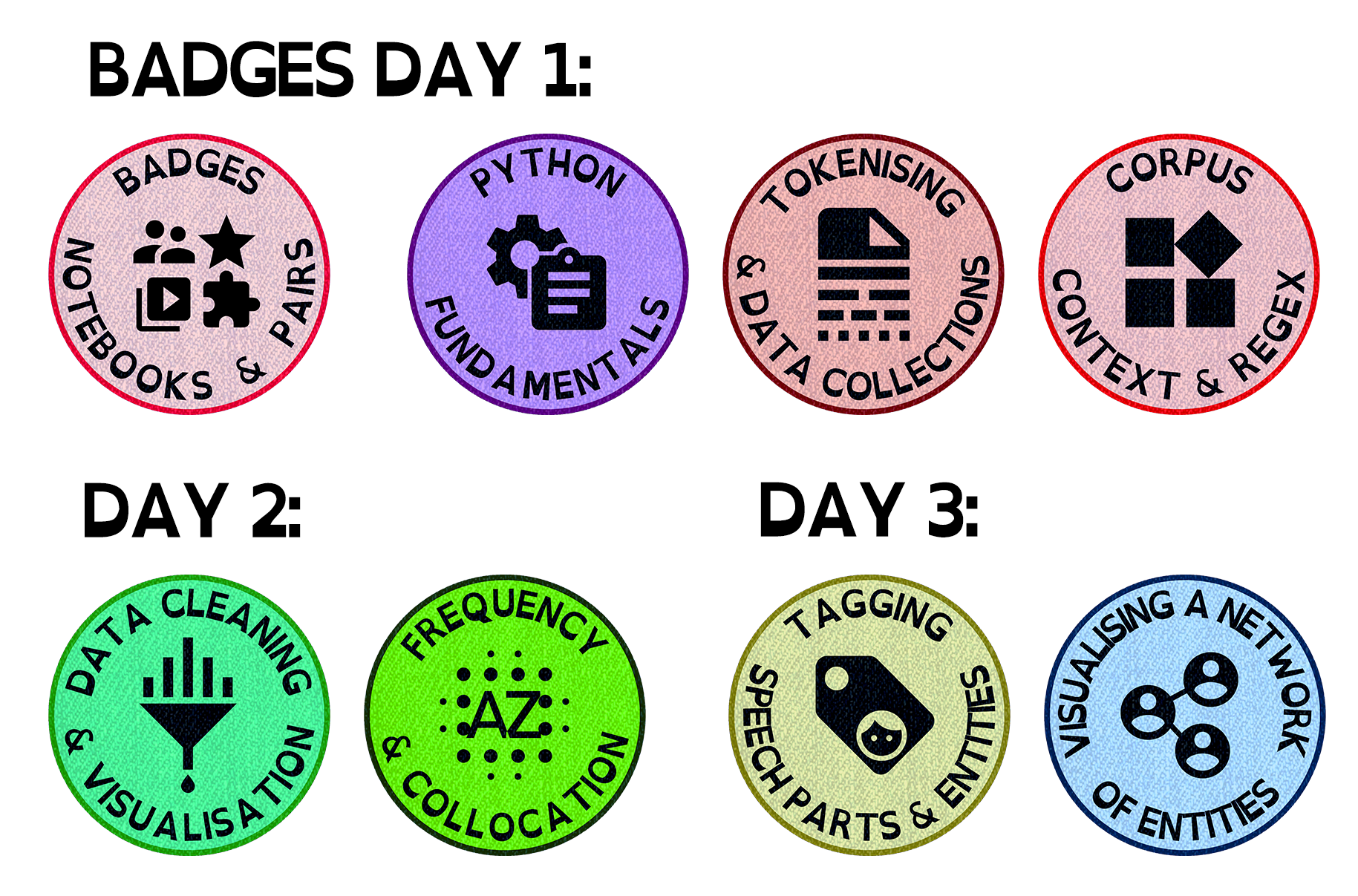}
  \caption{The badges used in our TDM course.  We created them using Android Material Design Icons which are open source under Apache License 2.0.}
  \label{fig:badges}
\end{figure}

The main change we made to our first pilot, without altering the course content, is that we restructured the course material into teaching with digital badges \cite{gibson2015digital,muilenburg2016digital} which are used in gamification of education \cite{dicheva2015gamification,ostashewski2015history}. The principles that guided us were: flexibility, compartmentalisation and empowering the learner. Each badge is built around a Threshold concept \cite{land2005threshold}, a core step or skill (a `eureka' moment) that opens the doors to further learning. Using a clear name and symbol, each badge signposts students' takeaways and how it fits within the top level learning journey (see Figure~\ref{fig:badges}). 

The macro-structure in which badges form our course is complemented by a micro-structure of each badge: background theory and instructional content, code-along videos, notebooks with worked examples, exercises of increasing difficulty, relentless feedback, pair work and mini coding problems (with solutions). Badges build on top of each other, forming branches and enabling optional, further learning. Additionally, the modular micro-structure, enables easier switching between platforms or teaching modes (e.g.~videos versus slides) and multiplies the benefits of improvements. Badges proved to be a promising format for delivering teaching of this course, especially in times of change, disruption and pivoting.

We wanted to give us and our course participants more flexibility, so we recorded all of the short lectures presented at the start of each badge and situated before each coding session in the course. This allowed students to come back to the recorded lecture materials later-on.  It also gave us more flexibility answering questions in the chat, solving technical issues in the background and discussing the running of a given badge in a teaching team break-out room while participants were watching the video lecture.

\section{Feedback}\label{section:feedback}

\subsection{Relentless Feedback}

\begin{figure*}[h!]
\centering
\begin{tikzpicture}
\node[inner sep=0pt] (fig) at (5,1)
    {\includegraphics[width=.65\textwidth]{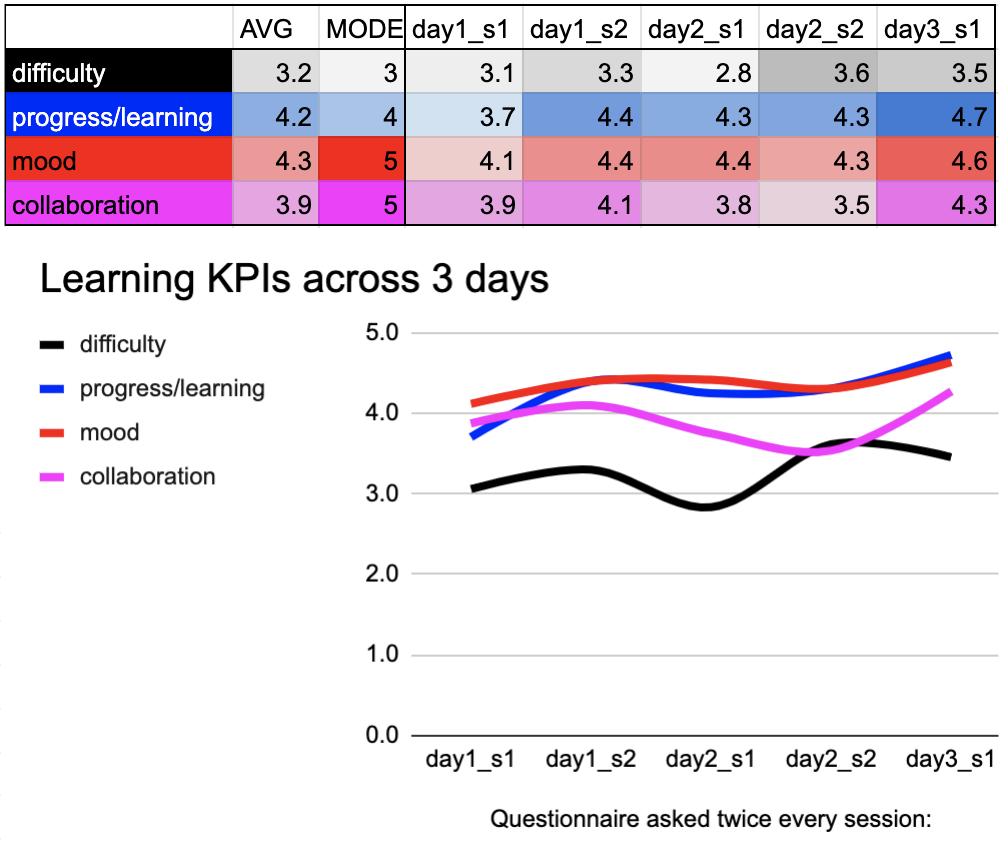}
};
\node[text width=10cm, inner sep=0pt] at (5,-6) {\textit{“Fantastic! The pair learning is excellent! Jupiter notebooks are a great tool. The real time interactivity is super rewarding.”}};    
\end{tikzpicture}
  \vspace{0.5cm}
  \caption{Feedback analysis for all surveys over the course of the boot camp and a quote from one student (with permission to share). We asked students to record difficulty of the course, their progress and learning, their mood and how they felt about their collaboration in pairs as key performance indicators (KPIs) throughout the course.}
  \label{fig:survey}
\end{figure*}

In both pilots we collected relentless feedback. This feedback loop helped us to address questions raised and go over things that were unclear.  We found it was really important to be flexible and adapt to what the students wanted. The twice-a-session mini-feedback form was really helpful for that and we made it very clear which parts of the course on day 2 and 3 were in response to participants' feedback (see feedback analysis in Figure~\ref{fig:survey}).

For example, a comments we received in the first pilot was that the students would prefer a quick recap of the previous session, which we then started doing and was a great way to link sessions and get the course material fresh in everyone’s minds.  Given the feedback, we also worked through the first section of a notebook together, so everyone had a clear idea of what to do.

The relentless feedback and our response is one of the reasons we believe we had such a high participant retention rate which we were very pleased about.  The pilots was free of charge, non-compulsory and ran over three afternoons. At least two thirds of the students who joined at the start of the week completed the last session on Friday.

We received constructive criticism but overall had very positive feedback on the course which, especially after the first pilot, made us feel very motivated having just had completed teaching our first online course. One participant thought it was “Fantastic!” in our final feedback survey. Another wrote “The pair learning is excellent! Jupiter [sic] notebooks are a great tool. The real-time interactivity is super rewarding.” Others reported that the lecturers and the “humour and playfulness of the examples” made the course “really great, especially for someone completely new to coding.” Yet another person commented that they would use the skills they learned in gathering data for their undergraduate dissertation about their research project.

\begin{figure*}[ht!]
\frame{\includegraphics[width=\textwidth]{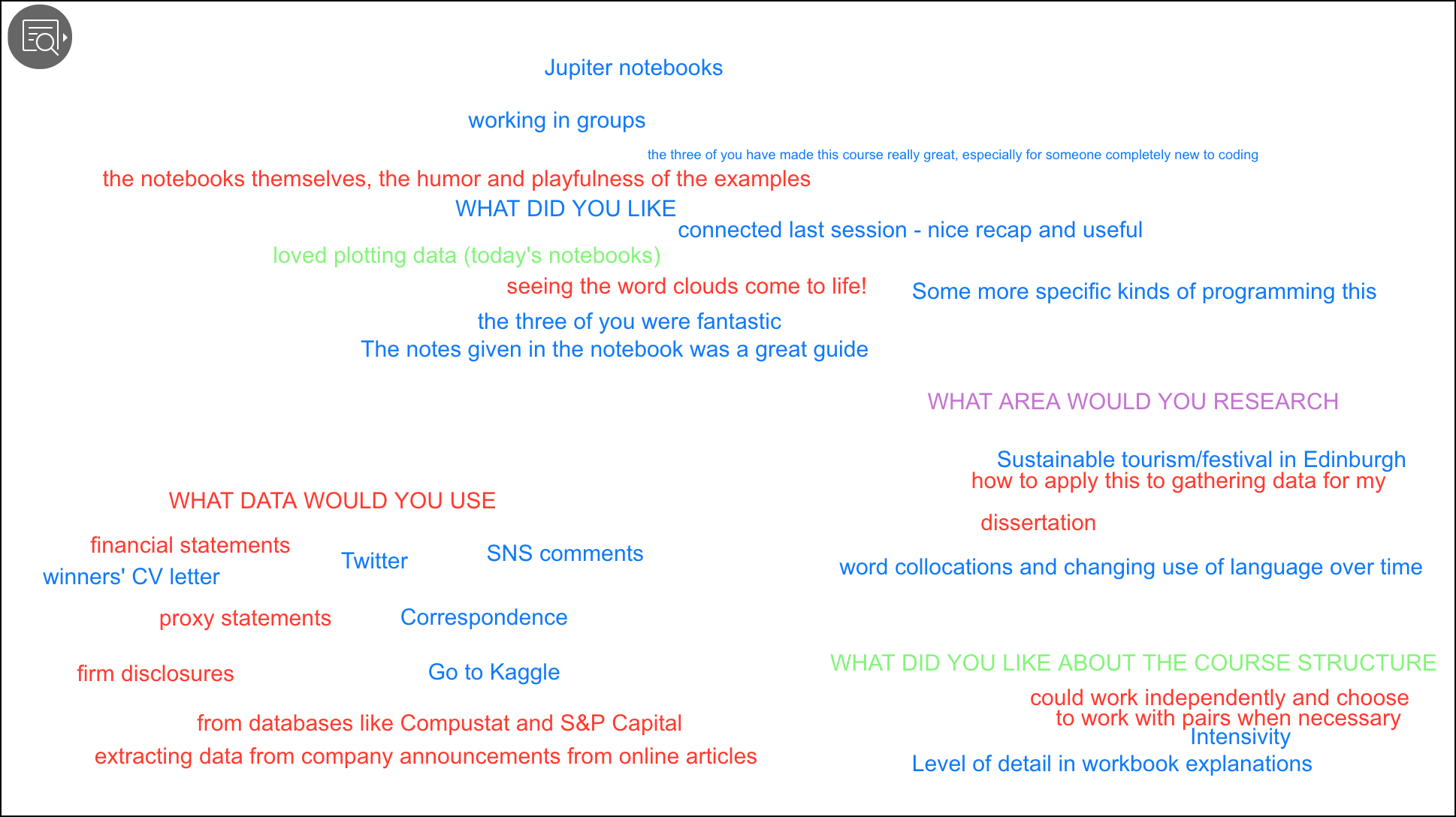}}
  \caption{Whiteboard with feedback generated by students in the course.}
  \label{fig:feedback}
\end{figure*}

\subsection{Detailed Student Feedback}\label{section:feedback-maria}
The following account is a more detailed reaction to our course provided by one of the student who participated in the first pilot of the TDM course and who we include as an author on this paper:

\textit{I was one of the mature students on the first pilot of the TDM Workshop – an academic myself with quantitative methods and coding experience in Stata and MatLab but not in Python, nor any previous experience with text mining or natural language processing.}

\textit{I appreciated the feedback requests at the end of each session via Microsoft Office forms and the immediate showcasing of the results for the whole class. Whenever there were bandwidth issues, the teaching team coordinated instantaneously and took over from each other.}

\textit{The part of the course that taught me the most were the pair breakout rooms where we worked through computational Jupyter notebooks. The annotation of the exercises was invaluable, as were the videos showing one of the instructors working through a notebook themselves and importantly running into an error and explaining how we use the error message as guidance to fix the code. During the breakout sessions having the three instructors drop in and answer any questions was an excellent balance of allowing the students independence while also feeling supported. Working with different partners every time was also very valuable. When taking an active role and talking through the lines of code and my understanding of the outcome, I was able to check in with my partner and be exposed to their style and approach to learning. Similarly, when taking the passive role and witnessing their way to working through a computational notebook, I could take away ideas of how to explain my thinking and understanding of the code in different ways.}

\textit{The distribution of new material via GitHub was very efficient. The interactions via the virtual whiteboard created playfulness and joy in the learning process. Although I did not participate in the second pilot and was not exposed to the Badges, I see them as another element of enhancing the playfulness of the process.}

\textit{The TDM Workshop I participated in took place relatively early in the pandemic before "Zoom fatigue" had set it and participants were excited to engage. A year later, full-time students appear to have become more resistant to engaging in voice and/or visual participation.}

\textit{There were some points that required improvement, for example typos in the annotation of the computational notebooks or some time being eaten up by technical troubleshooting. However, even these created an atmosphere of immediacy, flexibility and a sense of "We are all in this together".}

\textit{Overall, I benefited immensely from taking the first pilot. Not only do I now have an idea of text mining tools and how to use them but I was also inspired by and adopted the computational notebooks in my own teaching of Investments in the Autumn of 2020. I also implemented regular feedback, which I felt provided the element of playfulness and joy, in an even more interactive platform with gifs, wordclouds and animations (using Mentimeter\footnote{\url{https://www.mentimeter.com}}).}

\section{Lessons Learned}\label{section:lessons}

Despite on-the-whole positive comments, we still found teaching in an online environment quite odd. We felt that we lost the sense of whether the students were engaged, learning and enjoying the experience because most participants had their cameras switched off so we could not see their faces or body language. The feedback did help, even simply asking students to ‘raise your hand if you can hear me’, but it still remains odd to us to talk to a blank screen without seeing everyone.

We did not get everything right. The technology did not always work but luckily one teaching team member is quite experienced in fixing software-related issues.  We would have struggled without it.  Initially we also did not give enough thought to accessibility; we just assumed the software would deal with that - it did not. We learned that we have to ask all students before the course if they might have issues in accessing course materials or video calls and make time to deal with any technical issues that could arise as a result. 

We learned that students can be shy when it comes to talking to each other and putting on their webcams. We found ice breaker questions upfront, can be answered playfully on a whiteboard, very helpful for putting students at ease and have some fun. We used some simple things that made a lot of difference. We played music in the room before the class so when students joined, they knew we were there and that their speakers were on. We made extensive use of the virtual whiteboard to gather anonymous feedback really fast in addition to frequent short surveys (see Figure~\ref{fig:feedback}). We also included questions in the notebooks that buddies had to work on together to encourage discussion.  The notebooks contained essential TDM coding tasks and more complex tasks for the curious. This allowed some students to extend their learning without others feeling they were left behind.

We also found that the amount of content that we could cover grew as the course went on. There were initial issues with the technology which needed fixing and as we were all getting used to the new way of teaching.  The conversation also became more natural as time went on. At first it was quite odd to drop into the break-out rooms but by the second session this became easier and we were all chatting a lot more. The majority of students really liked the pair programming, they liked the flexibility and the content. They really felt they were part of the course in a way that is not always experienced online. 

As instructors, we found that teaching in this way, switching between modes, lecturing, answering the chat, live coding and responding to issues is really cognitively challenging. It is hard work and cannot easily be done by one individual. The technologies we use are complex and can fail but they are for the most part intuitive and provide a wide range of ways to teach and interact. We learned that online teaching is exhausting but done right it can still be really rewarding. We all enjoyed the interactions and felt part of a little community. After the course we did a debriefing and each wrote down three things we liked about the course and something we wished we could have achieved (see page Appendix on \pageref{sec:starsandwishes}).

We, the TDM course teachers on this paper, have, in the same way as the author who participated in the course, benefited immensely from what we learned through these pilots before delving into our online  teaching in the first term of 2020/21.

\section{Summary and Future Work}\label{section:summary-and-future}

In this paper, we have reflected of how we ported a TDM course online as a result of the global pandemic caused by COVID-19. We described the content of the course and how we adapted it over two pilot runs.  We particularly found different features of Blackboard Collaborate useful for teaching, especially the use of a virtual whiteboard and dividing the class up into break-out rooms.  Students responded positively to learning in pairs and to course materials broken down into digital badges. Finally, the relentless feedback we collected throughout each session and after the course helped us as teachers to improve the course and how we teach it. To make a course like this a good learning experience, it is really important to build community and get students to talk not just to the teachers but to each other as they would in a classroom.

Being caught in lockdown encouraged us to innovate, and our experience demonstrates what is possible to achieve virtually despite the limitations. Experiencing the learning in a classroom is difficult to replicate online, however, we are confident that these types of virtual environments will play a role in education beyond this pandemic, to complement and enhance traditional learning.

Going forward we would like to experiment with teaching this course in different ways: asynchronously to students joining from different time zones, to much larger groups to understand where the limits are in terms of number of participants given staff capacity, or in a writer-retreat type setup where the instructors touch base with students several times during the day.  We will also look at how this course can be pivoted back to on-campus teaching for students who can join in person and once the current pandemic slows down, lockdown restrictions are relaxed and on-campus teaching resumes.  We are pleased to announce that this course will be part of the post-graduate programme taught at EFI.

\section*{Acknowledgements}
We thank Professor Laura Cram for supporting us in the long-term development of this course as part of EFI's PGT training programme.  Our course was built on and expands material developed for a Library Carpentries course\footnote{\url{http://librarycarpentry.org/lc-tdm/}} with the support of the Centre for Data, Culture and Society.\footnote{\url{https://www.cdcs.ed.ac.uk}}  We would also like to thank Siobhan Dunn and her colleagues at EFI for managing the registration for our courses and Marco Rossi at the University of Edinburgh Business School's Student Development Team for offering the course to their students.

\bibliography{anthology,custom}

\begin{thebibliography}{24}
\expandafter\ifx\csname natexlab\endcsname\relax\def\natexlab#1{#1}\fi

\bibitem[{Agarwal(2013)}]{agarwal-2013-teaching}
Apoorv Agarwal. 2013.
\newblock \href {https://www.aclweb.org/anthology/W13-3412} {Teaching the
  basics of {NLP} and {ML} in an introductory course to information science}.
\newblock In \emph{Proceedings of the Fourth Workshop on Teaching {NLP} and
  {CL}}, pages 77--84, Sofia, Bulgaria. Association for Computational
  Linguistics.

\bibitem[{Alex et~al.(2019{\natexlab{a}})Alex, Grover, Tobin, and
  Oberlander}]{alex2019geoparsing}
Beatrice Alex, Claire Grover, Richard Tobin, and Jon Oberlander.
  2019{\natexlab{a}}.
\newblock \href {https://doi.org/10.1007/s10579-019-09443-x} {Geoparsing
  historical and contemporary literary text set in the {City of Edinburgh}}.
\newblock \emph{Language Resources and Evaluation}, 53(4):651--675.

\bibitem[{Alex et~al.(2019{\natexlab{b}})Alex, Grover, Tobin, Sudlow, Mair, and
  Whiteley}]{alex2019text}
Beatrice Alex, Claire Grover, Richard Tobin, Cathie Sudlow, Grant Mair, and
  William Whiteley. 2019{\natexlab{b}}.
\newblock \href {https://doi.org/10.1186/s13326-019-0211-7} {Text mining brain
  imaging reports}.
\newblock \emph{Journal of Biomedical Semantics}, 10(1):1--11.

\bibitem[{Bayne(2015)}]{bayne2015teacherbot}
Sian Bayne. 2015.
\newblock \href {https://doi.org/10.1080/13562517.2015.1020783} {Teacherbot:
  interventions in automated teaching}.
\newblock \emph{Teaching in Higher Education}, 20(4):455--467.

\bibitem[{Bird et~al.(2005)Bird, Klein, and Loper}]{bird2005nltk}
Steven Bird, Ewan Klein, and Edward Loper. 2005.
\newblock \href
  {https://cogsci.ucsd.edu/~rik/courses/readings/bird05-NLTK-intro.pdf} {{NLTK
  tutorial: Introduction to natural language processing}}.
\newblock \emph{Creative Commons Attribution}.

\bibitem[{Bird et~al.(2009)Bird, Klein, and Loper}]{bird2009natural}
Steven Bird, Ewan Klein, and Edward Loper. 2009.
\newblock \emph{{Natural language processing with Python: Analyzing text with
  the natural language toolkit}}.
\newblock " O'Reilly Media, Inc.".

\bibitem[{Clifford et~al.(2016)Clifford, Alex, Coates, Klein, and
  Watson}]{clifford2016geoparsing}
Jim Clifford, Beatrice Alex, Colin~M Coates, Ewan Klein, and Andrew Watson.
  2016.
\newblock \href {https://doi.org/10.1080/01615440.2015.1116419} {Geoparsing
  history: Locating commodities in ten million pages of nineteenth-century
  sources}.
\newblock \emph{Historical Methods: A Journal of Quantitative and
  Interdisciplinary History}, 49(3):115--131.

\bibitem[{Dicheva et~al.(2015)Dicheva, Dichev, Agre, and
  Angelova}]{dicheva2015gamification}
Darina Dicheva, Christo Dichev, Gennady Agre, and Galia Angelova. 2015.
\newblock \href {https://www.jstor.org/stable/jeductechsoci.18.3.75}
  {Gamification in education: A systematic mapping study}.
\newblock \emph{Journal of Educational Technology \& Society}, 18(3):75--88.

\bibitem[{Fawns et~al.(2019)Fawns, Aitken, and Jones}]{fawns2019online}
Tim Fawns, Gill Aitken, and Derek Jones. 2019.
\newblock \href {https://doi.org/10.1007/s42438-019-00048-9} {Online learning
  as embodied, socially meaningful experience}.
\newblock \emph{Postdigital Science and Education}, 1(2):293--297.

\bibitem[{Gibson et~al.(2015)Gibson, Ostashewski, Flintoff, Grant, and
  Knight}]{gibson2015digital}
David Gibson, Nathaniel Ostashewski, Kim Flintoff, Sheryl Grant, and Erin
  Knight. 2015.
\newblock \href {https://doi.org/10.1007/s10639-013-9291-7} {Digital badges in
  education}.
\newblock \emph{Education and Information Technologies}, 20(2):403--410.

\bibitem[{Gorinski et~al.(2019)Gorinski, Wu, Grover, Tobin, Talbot, Whalley,
  Sudlow, Whiteley, and Alex}]{gorinski2019named}
Philip~John Gorinski, Honghan Wu, Claire Grover, Richard Tobin, Conn Talbot,
  Heather Whalley, Cathie Sudlow, William Whiteley, and Beatrice Alex. 2019.
\newblock \href {http://arxiv.org/abs/1903.03985} {Named entity recognition for
  electronic health records: A comparison of rule-based and machine learning
  approaches}.
\newblock ArXiv.

\bibitem[{Hanks et~al.(2011)Hanks, Fitzgerald, McCauley, Murphy, and
  Zander}]{hanks2011pair}
Brian Hanks, Sue Fitzgerald, Ren{\'e}e McCauley, Laurie Murphy, and Carol
  Zander. 2011.
\newblock \href {https://doi.org/10.1080/08993408.2011.579808} {Pair
  programming in education: A literature review}.
\newblock \emph{Computer Science Education}, 21(2):135--173.

\bibitem[{Hearst(2005)}]{hearst2005teaching}
Marti~A Hearst. 2005.
\newblock \href {https://www.aclweb.org/anthology/W05-0101/} {Teaching applied
  natural language processing: Triumphs and tribulations}.
\newblock In \emph{Proceedings of the Second ACL Workshop on Effective Tools
  and Methodologies for Teaching NLP and CL}, pages 1--8.

\bibitem[{Hovy(2020)}]{hovy2020text}
Dirk Hovy. 2020.
\newblock \href {https://doi.org/10.1017/9781108873352} {\emph{Text Analysis in
  Python for Social Scientists: Discovery and Exploration}}.
\newblock Cambridge University Press.

\bibitem[{Jurgens and Li(2018)}]{jurgens2018look}
David Jurgens and Lucy Li. 2018.
\newblock \href
  {https://medium.com/@jurgens_24580/a-look-inside-the-the-pedagogy-of-natural-language-processing-9c5f4bdcf2a0}
  {{A Look Inside the Pedagogy of Natural Language Processing}}.
\newblock 25/09/2018. Accessed on 23/04/2021.

\bibitem[{Land et~al.(2005)Land, Cousin, Meyer, and Davies}]{land2005threshold}
Ray Land, Glynis Cousin, Jan~HF Meyer, and Peter Davies. 2005.
\newblock \href
  {https://citeseerx.ist.psu.edu/viewdoc/download?doi=10.1.1.644.7648&rep=rep1&type=pdf}
  {Threshold concepts and troublesome knowledge (3): implications for course
  design and evaluation}.
\newblock \emph{Improving student learning diversity and inclusivity},
  4:53--64.

\bibitem[{Llewellyn et~al.(2015)Llewellyn, Grover, Alex, Oberlander, and
  Tobin}]{llewellyn2015extracting}
Clare Llewellyn, Claire Grover, Beatrice Alex, Jon Oberlander, and Richard
  Tobin. 2015.
\newblock \href
  {https://link.springer.com/chapter/10.1007/978-3-319-24592-8_36} {{Extracting
  a topic specific dataset from a Twitter archive}}.
\newblock In \emph{International Conference on Theory and Practice of Digital
  Libraries}, pages 364--367. Springer.

\bibitem[{McGillivray et~al.(2020)McGillivray, Alex, Ames, Armstrong, Beavan,
  Ciula, Colavizza, Cummings, De~Roure, Farquhar
  et~al.}]{mcgillivray2020challenges}
Barbara McGillivray, Beatrice Alex, Sarah Ames, Guyda Armstrong, David Beavan,
  Arianna Ciula, Giovanni Colavizza, James Cummings, David De~Roure, Adam
  Farquhar, et~al. 2020.
\newblock \href {dx.doi.org/10.6084/m9.figshare.12732164} {{The challenges and
  prospects of the intersection of humanities and data science: A white paper
  from The Alan Turing Institute}}.

\bibitem[{Muilenburg and Berge(2016)}]{muilenburg2016digital}
Lin~Y Muilenburg and Zane~L Berge. 2016.
\newblock \href {https://doi.org/10.4324/9781315718569} {\emph{Digital badges
  in education: Trends, issues, and cases}}.
\newblock Routledge.

\bibitem[{of~Scotland(2019)}]{nls-mhbi2019}
National~Library of~Scotland. 2019.
\newblock \href {https://doi.org/10.34812/2w0t-3f08} {{A Medical History of
  British India}}.
\newblock Accessed on 23/04/2021.

\bibitem[{Ostashewski and Reid(2015)}]{ostashewski2015history}
Nathaniel Ostashewski and Doug Reid. 2015.
\newblock \href {https://doi.org/10.1007/978-3-319-10208-5_10} {A history and
  frameworks of digital badges in education}.
\newblock In \emph{Gamification in education and business}, pages 187--200.
  Springer.

\bibitem[{Ross et~al.(2013)Ross, Gallagher, and Macleod}]{ross2013making}
Jen Ross, Michael~Sean Gallagher, and Hamish Macleod. 2013.
\newblock \href {https://doi.org/10.19173/irrodl.v14i4.1545} {Making distance
  visible: Assembling nearness in an online distance learning programme}.
\newblock \emph{International Review of Research in Open and Distributed
  Learning}, 14(4):51--67.

\bibitem[{Williams(2006)}]{williams2006debunking}
Laurie Williams. 2006.
\newblock \href {https://doi.org/10.1109/MC.2006.160} {Debunking the nerd
  stereotype with pair programming}.
\newblock \emph{Computer}, 39(5):83--85.

\bibitem[{Williams et~al.(2000)Williams, Kessler, Cunningham, and
  Jeffries}]{williams2000strengthening}
Laurie Williams, Robert~R Kessler, Ward Cunningham, and Ron Jeffries. 2000.
\newblock \href {https://doi.org/10.1109/52.854064} {Strengthening the case for
  pair programming}.
\newblock \emph{IEEE software}, 17(4):19--25.

\end{thebibliography}
\bibliographystyle{acl_natbib}

\clearpage

\appendix
\section{Appendix: Three Stars and a Wish}\label{sec:starsandwishes}
\begin{figure}[!h]
\begin{tikzpicture}
\calloutquote[author=,width=9cm,position={(0,0)},fill=blue!20,inner sep=10pt,rounded corners]{
Clare: 
\begin{itemize}[label={},leftmargin=*]
    \item $\bigstar$ I liked that we were all willing to try anything and go beyond our comfort zone and fail
    \item $\bigstar$ I liked the way we naturally supported each other and took different roles, and swapped those roles
    \item $\bigstar$ I enjoyed learning the technology – something I thought I’d hate!
    \item 
    \sep \textit{I wish we could find a way to make the students more interactive, chat, turn their cameras on.}
\end{itemize}
}
\node[inner sep=0pt] (russell) at (7.5,0)
    {\roundpic[xshift=-0.1cm,yshift=-.5cm]{5.8cm}{6cm}{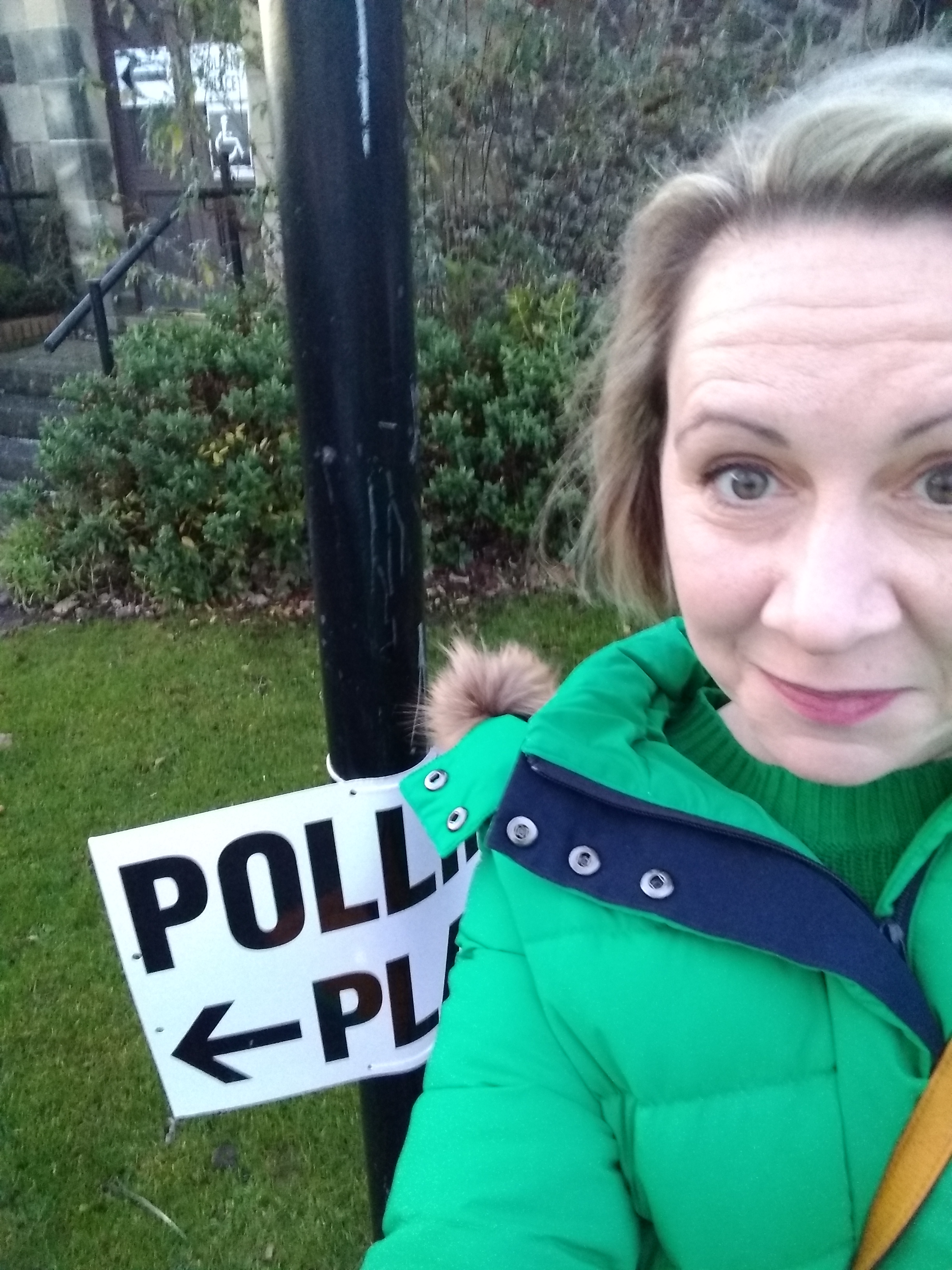}};
\end{tikzpicture} 

~

~

\begin{tikzpicture}
\calloutquote[author=,width=9.5cm,position={(0,0)},fill=blue!20,inner sep=10pt,rounded corners]{
Pawel:
\begin{itemize}[label={},leftmargin=*]
    \item $\bigstar$ Relentless feedback: asking 30s pulse-checking questionnaires twice a day; we used Office Forms.
    \item $\bigstar$ Staff room: easy and persisting internal comms channel within the course development team; we used Teams.
    \item $\bigstar$ Poetic licence: the person who created the final version of the notes was allowed to make adjustments to the content, so they fit the format.
    \item 
    \sep \textit{Our greatest ally was the tech that just worked. I wish we could also provide core hybrid experience of being in the same room.}
\end{itemize}
}
\node[inner sep=0pt] (russell) at (-7.5,0)
    {\roundpic[xshift=0cm,yshift=-.5cm]{5.8cm}{6cm}{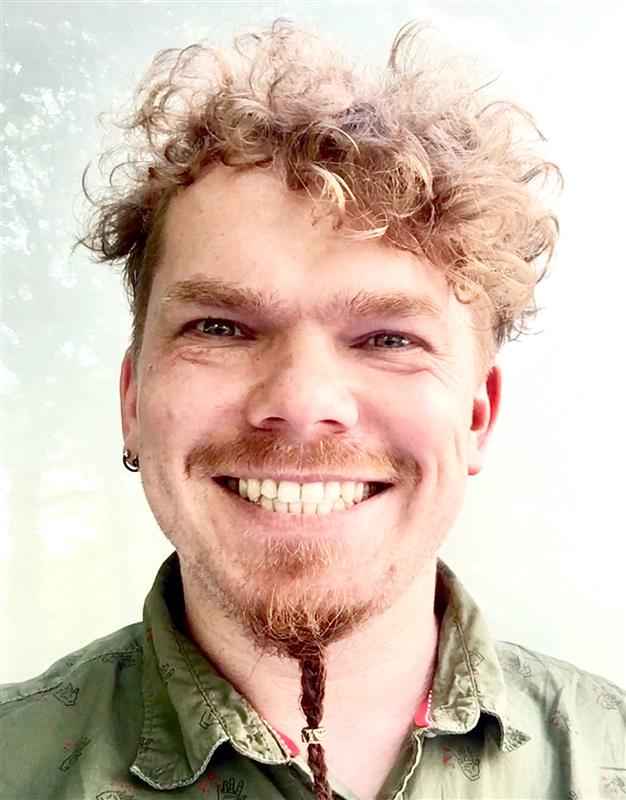}};
\end{tikzpicture}

~

\begin{tikzpicture}
\calloutquote[author=,width=9cm,position={(0,0)},fill=blue!20,inner sep=10pt,rounded corners]{
Beatrice:
\begin{itemize}[label={},leftmargin=*]
    \item $\bigstar$ Great team of people, great combination of skills
    \item $\bigstar$ Brilliant to see feedback from students coming in
    \item $\bigstar$ I learned a lot about teaching online
    \item \sep \textit{I wish I could witness the learning better online.}
\end{itemize}
}
\node[inner sep=0pt] (russell) at (7.5,0)
    {\roundpic[xshift=-0.5cm,yshift=-.5cm]{5.8cm}{7cm}{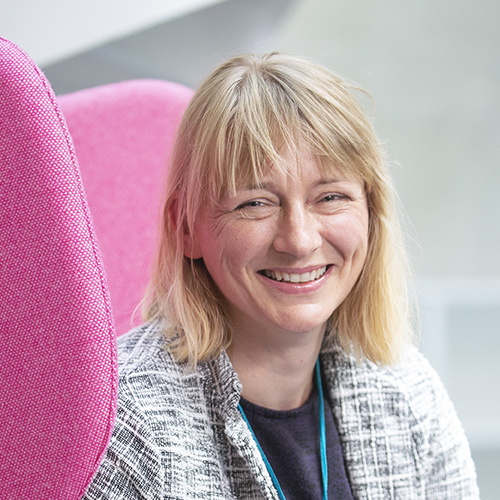}};
\end{tikzpicture}

\end{figure}

\end{document}